\documentclass[prd,twocolumn,showpacs,aps,superscriptaddress]{revtex4}
\usepackage{graphicx}
\usepackage{latexsym}
\usepackage{amssymb}

\begin{document}

\title{Remnants of dark matter clumps}

\author{Veniamin Berezinsky}
 \email{berezinsky@lngs.infn.it}
 \affiliation{INFN, Laboratori Nazionali del Gran Sasso, I-67010
  Assergi (AQ), Italy}
 \affiliation{Institute for Nuclear Research of the Russian Academy of
 Sciences, Moscow, Russia}
\author{Vyacheslav Dokuchaev}
 \email{dokuchaev@lngs.infn.it}
\author{Yury Eroshenko}
 \email{erosh@inr.npd.ac.ru}
 \affiliation{Institute for Nuclear Research of the Russian Academy of
 Sciences, Moscow, Russia}

\date{\today}

\begin{abstract}
What happened to the central cores of tidally destructed dark
matter clumps in the Galactic halo? We  calculate the probability
of surviving of the remnants of dark matter clumps in the Galaxy
by modelling the tidal destruction of the small-scale clumps. It
is demonstrated that a substantial fraction of clump remnants may
survive through the tidal destruction during the lifetime of the
Galaxy if the radius of a core is rather small. The resulting mass
spectrum of survived clumps is extended down to the mass of the
core of the cosmologically produced clumps with a minimal mass.
Since the annihilation signal is dominated by the dense part of
the core, destruction of the outer part of the clump affects the
annihilation rate relatively weakly and the survived dense
remnants of tidally destructed clumps provide a large contribution
to the annihilation signal in the Galaxy. The uncertainties in
minimal clump mass resulting from the uncertainties in neutralino
models are discussed.
\end{abstract}

\maketitle

\section{Introduction}

According to current observations, about 30\% of the mass of the
Universe is in a form of cold dark matter (DM). The nature of DM
particles  is still  unknown. The cold DM component is
gravitationally unstable and is expected to form the
gravitationally bounded clumpy structures from the scale of the
superclusters of galaxies and  down to very small clumps of DM.
The large-scale DM structures  are observed as the galactic halos
and clusters of galaxies. They are also seen in numerical
simulations. Theoretical study of DM clumps are important  for
understanding the properties of DM particles because annihilation
of DM particles in small dense clumps may result in visible
signal. The DM clumps in the Galaxy can produce the bright spots
in the sky in the gamma or X-bands \cite{RolMoo00}. A local
annihilation rate is proportional to the square of the DM particle
number density. Thus, the annihilation signal from small clumps
can dominate over diffuse component of DM in the halo.

The cosmological formation and evolution of small-scale DM clumps
have been studied in numerous works
\cite{SchSchWid99,SchHof00,bde03,ZTSH,GreHofSch05,bde06,DieKuhMad06,
Green07,bertsh,AngZha07,bde07,GioPieTor07}. The minimum mass of
clumps (the cutoff of the mass spectrum), $M_{\rm min}$ is
determined by the collisional and collisionless damping processes
(see e.~g. \cite{GreHofSch05} and references therein). Recent
calculations \cite{bertsh} show that the cutoff mass is related to
the friction between DM particles and cosmic plasma similar to the
Silk damping. In the case of the Harrison-Zeldovich spectrum of
primordial fluctuations with CMB normalization, the first
small-scale DM clumps are formed at redshift $z\sim60$ (for
$2\sigma$ fluctuations) with a mean density
$7\times10^{-22}$~g~cm$^{-3}$, virial radius $6\times10^{-3}$~pc
and internal velocity dispersion $80$~cm~s$^{-1}$ respectively.
Only a very small fraction of these clumps survives the early
stage of tidal destruction during the hierarchial clustering
\cite{bde03}. Nevertheless, these survived clumps may provide the
major contribution to the annihilation signal in the Galaxy
\cite{bde03,bde06,KamKou08,PieBerBra07,AndKom06}. At a high
redshift, neutralinos, considered as DM particles, may cause the
efficient heating of the diffuse gas \cite{MyeNus07} due to
annihilation in the dense clumps.

One of the unresolved problem of DM clumps is a value of the
central density or core radius. Numerical simulations give a
nearly power density profile of DM clumps. Both the
Navarro-Frenk-White and Moore profiles give formally a divergent
density in the clump center. A theoretical modelling of the clump
formation \cite{ufn1} predicts a power-law profile of the internal
density of clumps
\begin{equation}
 \rho_{\rm int}(r)=
 \frac{3-\beta}{3}\,\bar\rho\left(\frac{r}{R}\right)^{-\beta},
 \label{rho}
\end{equation}
where $\bar\rho$ and $R$ are  the mean internal density and a
radius of clump, respectively,  $\beta\simeq1.8-2$ and $\rho_{\rm
int}(r)=0$ at $r>R$. A near isothermal power-law profile
(\ref{rho}) with $\beta\simeq2$ has been  recently obtained in
numerical simulations of small-scale clump formation
\cite{DieMooSta05}.

It must be noted that density profiles of small-scale DM clumps
and large-scale DM haloes may be different. The galactic halos are
well approximated by the Navarro-Frenk-White profile outside of
the central core where dynamical resolution of numerical
simulations becomes insufficient. Different physical mechanisms
are engaged for formation of a central core during the formation
and evolution of clumps. A theoretical estimation of the relative
core radius of a DM clump $x_c=R_c/R$ was obtained in \cite{ufn1}
from energy criterion, $x_c\equiv R_c/R\simeq\delta_{eq}^3$, where
$\delta_{eq}$ is a value of density fluctuation at the beginning
of a matter-dominated stage. A similar estimate for DM clumps with
the minimal mass $\sim10^{-6}M_\odot$ originated from $2\sigma$
fluctuation peaks gives $\delta_{eq}\simeq0.013$ and
$R_c/R\simeq1.8\times10^{-5}$, respectively. In \cite{bde03}, the
core radius  $x_c\simeq 0.3\nu^{-2}$ has been obtained, where
$\nu$ is a relative height of the fluctuation density peak in
units of dispersion at the time of energy-matter equality (see
also Section~\ref{distrib}).  This value is a result of  the
influence of tidal forces on the motion of DM particles in the
clump at the stage of formation. This estimate may be considered
as an upper limit for the core radius or as the break-scale in the
density profile, e.g., a characteristic scale in the
Navarro-Frenk-White profile. It could be that a real core radius,
where the density ceases to grow, is determined by the relaxation
of small-scale perturbations inside the forming clump
\cite{DorLuk}. Another mechanism for core formation arises in the
``meta-cold dark matter model'' due to late decay of cold thermal
relics into lighter nonrelativistic particles with low phase-space
density \cite{Kap05,StrKapBul07}.

Nowadays, numerical simulations have a rather low space resolution
in the central region of clumps to determine the core radius. The
only example with some indication to presence of a core with
radius $x_c\simeq10^{-2}$ is numerical simulation of small-scale
clump formation \cite{DieMooSta05}. The special numerical
simulations with a sufficiently high space resolution to reveal
the real core radius are very requested.

In this work, we consider the relative core radius $x_c=R_c/R$ of
DM clumps as a free parameter in the range $0.001-0.1$. We
investigate the dependence of the probability of clump survival in
the Galaxy on this parameter under the action of tidal forces from
galactic disk and stars. As a preferred, value we consider
$x_c\simeq10^{-2}$ in spite of the numerical simulations
\cite{DieMooSta05}. The corresponding annihilation rate of DM
particles is proportional to their squared number density, and
thus is very sensitive to the value of the core radius.

In our earlier works \cite{bde06,bde07}, we used a simplified
criterium for a tidal destruction of clump. Namely, we postulated
that the clump is destructed if a total tidal energy gain
$\sum(\Delta E)_j$ after several disk crossings (or collisions
with stars) becomes of order of initial binding energy of a clump
$|E|$, i.e.
\begin{equation}
 \sum\limits_j(\Delta E)_j\sim|E|,
 \label{prevcrit}
\end{equation}
where summation goes over the successive disk crossings (or
encounters with stars). This criterium is justified in the
cosmological context of the DM clump formation because both the
formation of the density profile of the clump and its tidal
heating proceed during the same time of nonlinear evolution of
density perturbation. For the Galaxy case, a more detailed
consideration is needed to describe a tidal destruction of DM
clumps by stars. An improved approach includes a gradual mass loss
of systems \cite{gnedin2,TayBab01,DieKuhMad}, in particular, by
small-scale DM clumps \cite{ZTSH,GoeGneMooDieSta07}.

In this work, we will describe a gradual mass loss of small-scale
DM clumps assuming that only the outer layers of clumps are
involved and influenced by the tidal stripping. Additionally, we
assume that inner layers of a clump are not affected by tidal
forces. In this approximation, we calculate a continuous
diminishing of the clump mass and radius during the successive
Galactic disk crossings and encounters with the stars. We accept
now for criterium of clump destruction the diminishing of the
radius of tidally stripped clump down to the core radius. An
effective time of mass loss for the DM clump remains nearly the
same as in our previous calculations \cite{bde07}. However, the
clump destruction time has now quite different physical meaning:
it provides now a characteristic time-scale for the diminishing of
the clump mass and size instead of the total clump destruction.
This means that small remnants of clumps may survive in the
Galaxy. Respectively, these remnants would be an additional source
of amplification of the DM annihilation signal in the Galaxy.

\section{Tidal destruction of clumps by disk}
\label{disksec}

The kinetic energy gain of a DM particle with respect to the
center of a clump after one crossing of the Galactic disk is
\cite{OstSpiChe}
\begin{equation}
 \delta E=\frac{4g_m^2(\Delta z)^2m}{v_{z,c}^2}A(a),
 \label{egain}
\end{equation}
where $m$ is a constituent DM particle mass, $\Delta z$ is a
vertical  distance (orthogonal to the disk plane) of a DM particle
with respect to the center of the clump, $v_{z,c}$ is a vertical
velocity of the clump with respect to the disk plane at the moment
of disk crossing, and $A(a)$ is the adiabatic correction factor. A
gravitational acceleration near the disk plane is
\begin{equation}
 g_m(r)=2\pi G\sigma_s(r),
 \label{diskacc}
\end{equation}
where we use an exponential model for a surface density of disk
\begin{equation}
 \sigma_s(r)=\frac{M_d}{2\pi r_0^2}\,e^{-r/r_0}
 \label{diskmass}
\end{equation}
with $M_d=8\times10^{10}M_\odot$, $r_0=4.5$~kpc.

The factor $A(a)$ in (\ref{egain}) describes the adiabatic
protection from slow tidal effects \cite{Wein1}. This adiabatic
correction, further on referred to as the Weinberg correction, is
defined as an additional factor $A(a)$ to the values of energy
gain in the momentum approximation. This factor satisfies the
following asymptotic conditions: $A(a)=1$ for $a\ll1$ and
$A(a)\ll1$ for $a\gg1$. In \cite{gnedin2}, the following fitting
formula was proposed:
\begin{equation}
 A(a)=(1+a^2)^{-3/2}.
 \label{acor}
\end{equation}
Here the adiabatic parameter $a=\omega\tau_d$, where $\omega$ is
an orbital frequency of DM particle in the clump, $\tau_d\simeq
H_d/v_{z,c}$ is an effective duration of gravitational tidal shock
produced by the disk with a half-thickness $H_d$. For tidal
interactions of clumps with stars in the bulge and the halo the
duration of the gravitational shock can be estimated as
$\tau_s\sim l/v_{\rm rel}$, where $l$ is an impact parameter and
$v_{\rm rel}$ is a relative velocity of a clump with respect to a
star.

As a representative example, we consider the isothermal internal
density profile of a DM clump
\begin{equation}
\rho_{\rm int}(r)=\frac{1}{4\pi}\frac{v_{\rm rot}^2}{Gr^2}
 \label{iso}
\end{equation}
with a cutoff at the virial radius $R$: $\rho(r)=0$ at $r>R$. A
corresponding mass profile of a clump is $M(r)=M_i(r/R)$, where
$M_i$ is an initial mass of a clump at the epoch of the Galaxy
formation. With this mass distribution, a circular velocity inside
a clump is independent of the radius, $v_{\rm
rot}=(GM(r)/r)^{1/2}=(GM_i/R)^{1/2}$. A gravitational potential
corresponding to the density profile (\ref{iso}) is
$\phi(r)=v_{\rm rot}^2[\log(r/R)-1]$. Let us define a
dimensionless energy of the DM particle $\varepsilon=E/(mv^2_{\rm
rot})$ and gravitational potential $\psi(r)=\phi(r)/v^2_{\rm
rot}=\ln(r/R)-1$. An internal density profile $\rho_{\rm int}(r)$
and the distribution function of DM particles in the clump $f_{\rm
cl}(\varepsilon)$ are related by the integral relation
\cite{Edd16}
\begin{equation}
 \rho_{\rm int}(r)=2^{5/2}\pi\int\limits_{\psi(r)}^{0}
 \sqrt{\varepsilon-\psi(r)}\,f_{\rm
 cl}(\varepsilon)\,d\varepsilon.
 \label{ferho}
\end{equation}
The corresponding isothermal distribution function is
\begin{equation}
 f_{\rm cl}(\varepsilon)\simeq
 \frac{v_{\rm rot}^2}{4\pi^{5/2}e^2GR^2}e^{-2\varepsilon}.
\end{equation}
Note what this distribution function provides only an approximate
representation of (\ref{iso}), far from the cutoff radius $R$.
Nevertheless, this approximation is enough for our estimates of
tidal destruction of the DM clumps.

By using the hypothesis of a tidal stripping of outer layers of a
DM clump, we see that a tidal energy gain $\delta\varepsilon$
causes the stripping of particles with energies in the range
$-\delta\varepsilon<\varepsilon<0$. A corresponding variation of
density at radius $r$ is
\begin{equation}
\delta \rho(r)=2^{5/2}\pi\int\limits_{-\delta\varepsilon}^{0}
 \sqrt{\varepsilon-\psi(r)}\,f_{\rm cl}(\varepsilon)\,d\varepsilon.
 \label{ferho2}
\end{equation}
In this equation, the tidal energy gain (\ref{egain}) by different
DM particles is averaged over angles, so as $\langle(\Delta
z)^2\rangle=r^2/3$. A resulting total mass loss by a DM clump
during one crossing of the Galactic disk is
\begin{equation}
\delta M=-4\pi\int_0^R r^2\delta\rho(r)\,dr.
\end{equation}
Let us specify the dimensionless quantities
\begin{equation}
 Q_d=\frac{g_m^2}{2\pi v_{z,c}^2G\bar\rho_i}, \quad
 S_d=\frac{4\pi}{3}G\bar\rho_i\tau^2_d,
 \label{eqqq}
\end{equation}
where $\bar\rho_i=3M_i/(4\pi R^3)$ is a initial mean density of
clump. For the most parts of clumps $Q_d\ll 1$ with a typical
value $Q_d\sim0.03$. In the limiting case $Q_d\ll 1$ and in the
absence of the adiabatic correction, $S_d=0$, the integrals
(\ref{ferho2}) can be calculated analytically. In a general case,
the fitting formula for the mass loss of a clump during one
passage through the Galactic disk is
\begin{equation}
 \left(\frac{\delta M}{M}\right)_d\simeq
 -0.13Q_d\exp\left(-1.58S_d^{1/2}\right).
 \label{mmq}
\end{equation}
Now, we calculate the tidal mass loss by clumps using a realistic
distribution of their orbits in the the halo. The method of
calculation is similar to the one used in \cite{bde07}, but
instead of the rough energetic criterium for a tidal clump
destruction (\ref{prevcrit}) we will assume now a gradual
decreasing of clump mass and size.

Let us choose some particular clump moving in the spherical halo
with an orbital ``inclination'' angle $\gamma$ between the normal
vectors of the disk plane and orbit plane. The orbit angular
velocity at a distance $r$ from the Galactic center is
$d\phi/dt=J/(mr^2)$, where $J$ is an orbital angular momentum of a
clump. A vertical velocity of a clump crossing the disk is
\begin{equation}
 v_{z,c}=\frac{J}{mr_s}\sin\gamma,
 \label{vzc}
\end{equation}
where $r_s$ is a radial distance of a crossing point from the
Galaxy center. There are two crossing points (with different
values of $r_s$) during an orbital period.

The standard Navarro-Frenk-White profile of the DM Galactic halo
is
\begin{equation}
 \rho_{\rm H}(r)=
 \frac{\rho_{0}}{\left(r/L\right)\left(1+r/L\right)^2},
 \label{halonfw}
\end{equation}
where $L=45$~kpc, $\rho_{0}=5\times10^6M_\odot$~kpc$^{-3}$. It
useful to introduce the dimensionless variables:
\begin{equation}
 x=\frac{r}{L}, \quad \tilde\rho_{\rm H}(x)=\frac{\rho_{\rm
 H}(r)}{\rho_0}, \quad y=\frac{J^2}{8\pi G\rho_0L^4M^2},
\end{equation}
\begin{equation}
 \varepsilon=\frac{E_{\rm orb}/M-\Phi_0}{4\pi G\rho_0L^2 }, \quad
 \psi=\frac{\Phi-\Phi_0}{4\pi G\rho_0L^2},
\end{equation}
where $\Phi_0=-4\pi G\rho_0L^2$, $E_{\rm orb}$ is a total orbital
energy of a clump. With these variables, the density profile of
the halo (\ref{halonfw}) is written as
\begin{equation}
 \tilde\rho_{\rm H}(x)=\frac{1}{x(1+x)^2}.
 \label{nfwhalox}
\end{equation}
A gravitational potential $\psi(x)$, corresponding to density
profile (\ref{nfwhalox}) is
\begin{equation}
 \psi(x)=1-\frac{\log(1+x)}{x}.
 \label{pot}
\end{equation}
An equation for orbital turning points, $\dot r^2=0$, for DM
clumps in the potential (\ref{pot}) is
\begin{equation}
 1-\frac{\log(1+x)}{x}+\frac{y}{x^2}=\varepsilon.
 \label{turn}
\end{equation}
From (\ref{turn}), one can find numerically the minimum $x_{\rm
min}$ and maximum $x_{\rm max}$ radial distance of a clump from
the Galactic center as a function of orbital energy $\varepsilon$
and square of angular momentum $y$. Denoting $p=\cos\theta$, where
$\theta$ is an angle between the radius-vector $\vec r$ and the
orbital velocity $\vec v$, we have
$y=(1-p^2)x^2[\varepsilon-\psi(x)]$. As we assumed above, the unit
vectors $\vec v/v$ are distributed isotropically at each point
$x$, and, therefore,  $p$ has a uniform distribution in the
interval $[-1,1]$.

The relation between the density profile $\tilde\rho_H(x)$ and the
distribution function is given by the same equation (\ref{ferho})
with an obvious substitution $f_{\rm cl}\Rightarrow
F(\varepsilon)$, where the distribution function $F(\varepsilon)$
for a halo profile (\ref{nfwhalox}) can be fitted as \cite{Wid00}
\begin{equation}
 F(\varepsilon)=F_1(1-\varepsilon)^{3/2}\varepsilon^{-5/2}
 \left[-\frac{\ln(1-\varepsilon)}{\varepsilon}\right]^qe^P.
 \label{fffrhofd}
\end{equation}
Here $F_1=9.1968\times10^{-2}$, $q=-2.7419$,
$P=\sum\limits_{i}p_i(1-\varepsilon)^i$,
$(p_1,p_2,p_3,p_4)=(0.3620, -0.5639, -0.0859, -0.4912)$. An
interval of time for motion from $x_{\rm min}$ to $x_{\rm max}$
and back is
\begin{equation}
 T_c(x,\varepsilon,p)=\frac{1}{\sqrt{2\pi G\rho_0}}
 \int\limits_{x_{\rm min}}^{x_{\rm max}}
 \frac{ds}{\sqrt{\varepsilon-\psi(s)-y/s^2}}.
 \label{pcint}
\end{equation}
An angle of orbital precession during the time $T_c/2$ is
\begin{equation}
 \tilde\phi=y^{1/2}\int\limits_{x_{\rm min}}^{x_{\rm max}}
 \frac{ds}{s^2\sqrt{\varepsilon-\psi(s)-y/s^2}}-\pi<0.
 \label{prec}
\end{equation}
Therefore, an orbital period is longer than $T_c$ and is given by
\begin{equation}
 T_t=T_c\left(1+\tilde\phi/\pi\right)^{-1}.
 \label{pcintt}
\end{equation}
Choosing a time interval $\Delta T$ much longer than a clump
orbital period $T_t$, but much shorter than the age of the Galaxy
$t_0$, i.e., $T_t\ll\Delta T\ll t_0$, we define an averaged rate
of mass loss by a selected clump under influence of tidal shocks
in successive disk crossings
\begin{equation}
\frac{1}{M}\left(\frac{dM}{dt}\right)_d\simeq\frac{1}{\Delta
T}\sum \left(\frac{\delta M}{M}\right)_d,
 \label{deriv2}
\end{equation}
where $(\delta M/M)_d$ is given by (\ref{mmq}) and summation goes
over all successive crossing points (odd and even) of a clump
orbit with the Galactic disk during the time interval $\Delta T$.
According to (\ref{diskacc}) and (\ref{vzc}) the $g_m$ and
$v_{z,c}$ both depend on the radius $x=r/L$. One simplification in
calculation of (\ref{deriv2}) follows from the fact that a
velocity of orbit precession is constant. For this reason the
points of successive odd crossings are separated by the same
angles $\tilde\phi$ from (\ref{prec}). The same is also true for
successive even crossings. Using this simplification, we transform
the summation in (\ref{deriv2}) to integration:
\begin{eqnarray}
 \frac{1}{\Delta T} \sum\left(\frac{\delta M}{M}\right)_d
 \simeq\frac{2}{T_t|\tilde\phi|}\int\limits_{x_{\rm min}}^{x_{\rm max}}
 \left(\frac{\delta M}{M}\right)_d\frac{d\phi}{dx}dx,
 \nonumber
\end{eqnarray}
where
\begin{equation}
 \frac{d\phi}{dx}= \frac{y^{1/2}}{x^2\sqrt{\varepsilon-\psi(x)-y/x^2}}
\end{equation}
is an equation for the clump orbit in the halo. The method
described will be used in the Section~\ref{secsurv} for the final
calculations.

\section{Tidal destruction of clumps by stars}
\label{starsec}

Now, we calculate the diminishing of a clump mass due to a tidal
heating by stars in the Galaxy by using the same hypothesis of the
the preferable stripping of the outer clump layers. During a
single close encounter of a DM clump with a star, the energy gain
of a constituent DM particle in the clump with respect to clump
center is \cite{bde06}
\begin{equation}
 \delta E=\frac{2G^2m_s^2m\Delta z^2}{v_{\rm rel}^2l^4},
 \label{egainstar}
\end{equation}
where $m_*$ is a star mass, $l$ is an impact parameter, $v_{\rm
rel}$ is a relative star velocity with respect to a clump, $\Delta
z=r\cos\psi$, $r$ is a radial distance of a DM particle from the
clump center, and $\psi$ is an angle between the directions from
the clump center to the DM particle and to the point of the
closest approach of a star. Using the same method as in
Sec.~\ref{disksec}, we calculate a relative mass loss by the clump
$(\delta M/M)_s$ during a single encounter with a star and obtain
the same fitting formula as (\ref{mmq}) but with substituting the
dimensionless parameters, $Q_d\Rightarrow Q_s$ and $S_d\Rightarrow
S_s$, where
\begin{equation}
 Q_s=\frac{Gm_*^2}{2\pi v_{\rm rel}^2l^4\bar\rho_i},
 \quad
 S_s=\frac{4\pi}{3}G\bar\rho_i\tau^2_s,
 \label{eqqq2}
\end{equation}
where $\tau_s\simeq l/v_{\rm rel}$.

A DM clump acquires the maximum energy gain during a single
encounter with a star when the impact parameter $l\sim R$. Using
the relation
\begin{equation}
 dt=\frac{1}{2\sqrt{2\pi G\rho_0}}
 \frac{dx}{\sqrt{\varepsilon-\psi(x)-y/x^2}},
 \label{dtdx}
\end{equation}
and integrating over all impact parameters $l>R$, we calculate an
averaged rate of mass loss by a clump during successive encounters
with stars
\begin{eqnarray}
 \label{mmqstar}
&& \frac{1}{M}\left(\frac{dM}{dt}\right)_s\simeq \\ &&
 \!\!\frac{1}{2T_t\sqrt{2\pi G\rho_0}} \int\limits_{R}^{\infty}\!2\pi
 l\,dl\!\int\limits_{x_{\rm min}}^{x_{\rm max}}\!\!
 \frac{ds\, n_*(s)v_{rel}}{\sqrt{\varepsilon-\psi(s)-y/s^2}}
 \left(\frac{\delta M}{M}\right)_{\!\!s}\!,  \nonumber
\end{eqnarray}
where $n_*(r)$ is a radial number density distribution of stars in
the bulge and halo. A DM clump moves through the medium with a
varying value of $n_*$ along the clump orbit. In contrast to the
case of the disk crossing, the precession of the clump orbit
during an orbital period does not influence the mass loss due to
encounters with stars. Additionally, the mass loss due to
encounters with stars is independent of the inclination of clump
orbits in the case of a spherically symmetric distribution of
stars in the bulge and halo.

Using the results of \cite{LauZylMez}, we approximate the radial
number density distribution of stars in the bulge in the radial
range $r=1-3$~kpc as
\begin{equation}
 n_{b,*}(r)=(\rho_b/m_*)\exp\left[ -(r/r_b)^{1.6}\right],
 \label{rhoe}
\end{equation}
where $\rho_b=8M_\odot/$pc$^3$ and $r_b=1$~kpc. A corresponding
number density distribution of the halo stars at $r>3$~kpc outside
the Galactic plane can be approximated as
\begin{equation}
 n_{h,*}(r)=(\rho_h/m_*) (r_{\odot}/r)^{3},
 \label{rhosh}
\end{equation}
where $m_*=0.4M_\odot$ and $r_{\odot}=8.5$~kpc. According to
\cite{Bell07}, in the region between $r=1$ and $40$~kpc a total
mass of stars is $4\times10^8M_\odot$ with a star density profile
$\propto r^{-3}$. These data correspond to
$\rho_h=1.4\times10^{-5}~M_\odot/$pc$^3$ in (\ref{rhosh}). We
neglect in our calculations the oblationes of the stellar halo
\cite{Bell07}.

\section{Surviving fraction of clumps}
\label{secsurv}

From Eq.~(\ref{egain}) it is seen that the tidal forces influence
mainly the outer part of the clump (where $\Delta z$ is rather
large). Further, we will use our basic assumption that only outer
layers of a clump undergo the tidal stripping, while the inner
parts of a clump are unaffected by tidal forces. Thus, we assume
that a clump mass $M=M(t)$ and radius $R=R(t)$ are both
diminishing in time due to the tidal stripping of outer layers,
but its internal density profile remains the same as given by
Eq.~(\ref{iso}), e.g., for the isothermal density profile
$M(t)\propto R(t)$ and $\bar\rho(t)\propto M(t)^{-2}$. Combining
together the rates of mass loss (\ref{deriv2}) and (\ref{mmqstar})
due to the tidal stripping of a clump by the disk and stars
respectively, we obtain the evolution equation for a clump mass:
\begin{equation}
 \frac{dM}{dt}=
 \left(\frac{dM}{dt}\right)_d+\left(\frac{dM}{dt}\right)_s
 \label{mmtmain}.
\end{equation}
In the following, we solve this equation numerically starting from
the time of Galaxy formation at $t_0-t_G$ up to  the present
moment $t_0$. In numerical calculations, it is convenient to use
the dimensionless variables: $t/t_0$ for time and $M/M_i$ for a
clump mass, where $M_i$ is an initial clump mass. The adiabatic
correction provides generally only a small effect. In the absence
of adiabatic correction or, equivalently, at $S_d=S_s=0$ the
evolution equation (\ref{mmtmain}) has a simple form
\begin{equation}
 \frac{d\mu}{dt}=-\frac{\mu}{t_s}-\frac{\mu^3}{t_d},
\end{equation}
where $\mu=M(t)/M_i$ and parameters $t_d$ and $t_s$ are
independent of $\mu$. The solution of this equation
\begin{equation}
 \mu^2(t_0)=\frac{2t_d}{(2t_d+t_s)\exp(2t_0/t_s)-t_s}
\end{equation}
represents a good approximation to numerical solution of
(\ref{mmtmain}) with the adiabatic correction taken into account.

The most important astrophysical manifestation of DM clumps is a
possible annihilation of constituent DM particles. The crucial
point is a dominance of the central core of a clump in
annihilation signal if clumps have a steep enough density profile.
Namely, annihilation of DM particles in a clump core will prevail
in a total annihilation rate in a single clump with a power-law
density profile (\ref{rho}) if $\beta>3/2$ and $x_c=R_c/R\ll1$.
More specifically, the quantity $\dot N\propto\int_{r_0}^{r}4\pi
r'^2dr'\rho_{\rm int}^2(r')$ practically does not depend on $r$,
if $r \gg r_0$. As a result, the annihilation luminosity of a DM
clump with approximately an isothermal density profile
($\beta\simeq 2$) will be nearly constant under influence of tidal
stripping until a clump radius diminishes to its core radius. In
other words, in the nowadays Galaxy the remnants of tidally
stripped clumps with $x_c<\mu(t_0)\ll1$, where $\mu(t)=M(t)/M_i$
and $t_0\simeq10^{10}$~yrs is the Galaxy age, obeys the evolution
equation (\ref{mmtmain}) and have the same annihilation luminosity
as their progenitors with $\mu=1$.

By using the evolution equation (\ref{mmtmain}), we now calculate
the probability $P$ of the survival of clump remnant during the
lifetime of the Galaxy. Let us choose some arbitrary point in the
halo with a radius-vector $\vec r$ and an angle $\alpha$ with a
polar axis of the Galactic disk. Only the clump orbits with an
inclination angle $\pi/2-\alpha<\gamma<\pi/2$ pass through this
point. A survival probability for clumps can be written now in the
following form
\begin{eqnarray}
 \label{sp1}
 P(x,\alpha)&=&\frac{4\pi\sqrt{2}}{\tilde\rho(x)\sin\alpha}
 \int\limits_{0}^{1}dp\int\limits_{0}^{\sin\alpha}d\cos\gamma \\
 &\times&\!\!\int\limits_{\psi(x)}^{1}\!d\varepsilon\,
 [\varepsilon-\psi(x)]^{1/2}F(\varepsilon){\rm \Theta}[\mu(t_0)-x_c].
 \nonumber
 \end{eqnarray}
In this equation, $\tilde\rho(x)$ is a density profile of the halo
from (\ref{nfwhalox}), $p=\cos\theta$, $\theta$ is an angle
between the radius-vector $\vec r$ and the orbital velocity of
clump, ${\rm \Theta}$ is the Heaviside function, $\psi(x)$ is the
halo gravitational potential from (\ref{nfwhalox}),
$F(\varepsilon)$ is a distribution function of clumps in the halo
from (\ref{fffrhofd}), $\mu(t_0)$ depends on all variables of the
integration, and $x_c=R_c/R$ is an initial value of the clump
core. The function $\mu(t_0)$ is calculated from numerical
solution of evolution equation (\ref{mmtmain}). If $\mu(t_0)>x_c$,
the clump remnant is survived through the tidal destruction by
both the disk and stars. The annihilation rate in this remnant
would be the same as in the initial clump. On the contrary, in the
opposite case, when $\mu(t_G)<x_c$, the clump is totally
destructed because (i) the core is not a dynamically separated
system and composed of particles with extended orbits, and because
(ii) a nearly homogeneous core is destructed easier than a similar
object with the same mass but with a near isothermal density
profile.

\begin{figure}[t]
\includegraphics[width=0.45\textwidth]{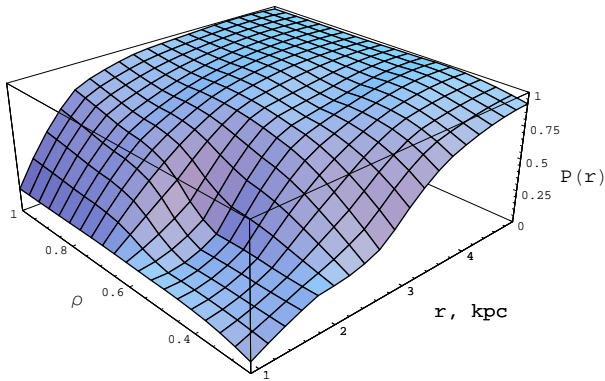}
\caption{The survival probability $P(r,\rho)$ plotted as a
function of distance from the Galactic center $r$ and a mean
internal clump density $\rho$ in the case $x_c=0.1$. It gives the
normalized fraction of DM clumps in the halo $P$ calculated from
(\ref{sp1}), which survives the tidal destruction by the stellar
disk and the halo stars.}
 \label{ani1f}
\end{figure}

\begin{figure}[t]
\includegraphics[width=0.45\textwidth]{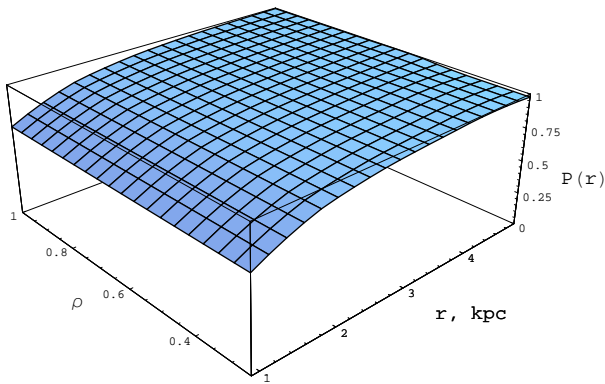}
\caption{The same as Fig.~\ref{ani1f}, but for the case
$x_c=0.05$.}
 \label{ani0f}
\end{figure}

We consider the small-scale DM clumps in the initial mass interval
$M_i=[10^{-6}M_\odot,1\,M_\odot]$ originated from the $2\sigma$
(i.e. $\nu=2$) peaks in the Harrison-Zeldovich perturbation
spectrum. A reason is as follows. The DM clumps originated from
initial density perturbations with $\nu<1$ were almost completely
destructed by tidal interactions during the early stage of
hierarchical clustering as it can be seen from the distribution
function of clumps (\ref{psiitog}) (see below). On the contrary,
the most dense DM clumps with $\nu>3$ are mostly survived the
stage of hierarchical clustering, but their number according to
(\ref{psiitog}) is exponentially falls with $\nu$ and is small.
For this reason, we will use the following approximation: the DM
clumps were originated on average from $\nu\simeq2$ peaks, and for
any given mass $M$ we do not consider the distribution of clumps
over their densities. In this approximation, the initial radius of
the clump $R_i$ depends only on the one parameter --- the initial
clump density, which also depends only on the initial clump mass
$M_i$.

The crucial result of numerical calculation of a survival
probability (\ref{sp1}) for clumps with $x_c\ll 0.05$ is that
$P(x,\alpha)\sim1$ everywhere. Even inside the bulge there are
clumps which flying through the bulge from external regions. These
means that clump remnants are mostly survived through the tidal
destruction in the Galaxy. A noticeable diminishing
$P(x,\alpha)<1$ near the center of Galaxy becomes apparent for
clumps with $x_c>0.05$. It is understandable because with
$x_c\to1$, we return to the previous criterium of tidal
destruction of clumps (\ref{prevcrit}) and to a corresponding
results for survival probability \cite{bde06,bde07}. The survival
probability $P(r,\alpha)$ numerically calculated from (\ref{sp1})
for the cases $x_c\sim0.05$ and $0.1$ is shown in the
Figs.~\ref{ani1f} and \ref{ani0f}. The dependence on $\alpha$ (an
angle between a radius-vector $\vec r$ and a polar axis of the
Galactic disk) is very weak as it was shown in \cite{bde07}. For
this reason, we present the results only for an intermediate value
$\alpha=\pi/4$. The density of clumps is normalized to the density
$7.3\times10^{-23}$~g~cm$^{-3}$ valid for clumps with mass
$M=10^{-6}M_\odot$ originated from $2\sigma$ density peaks in the
case of power-law index of primordial spectrum of perturbations
$n_p=1$.

It is worth to note that a tidal radius of a clump in the bulge is
\cite{King62}
\begin{equation}
 r_t^3=\frac{GM(r_t)}{\omega_p^2-d^2\phi/dl^2},
 \label{tid1}
\end{equation}
where $\omega_p\simeq[GM_b(l)/l^3]^{1/2}$ is an angular velocity
at the pericenter (we consider here a circular orbit for
simplicity), $M_b(l)$ is a mass profile of the bulge and $\phi(l)$
is a gravitational potential of the bulge. For the considered
small-scale clumps $r_t\ge0.2R_i$, and, therefore, the tidal
radius is not a crucial factor for destruction of clumps.

\section{Cosmological distribution function of clumps}
\label{distrib}

In this section, we provide calculations of a mass function for
the small-scale clumps in the Galactic halo by more transparent
method than in our previous works \cite{bde03,bde06}.

The first gravitationally bound objects in the Universe are the DM
clumps of minimum mass $M_{\rm min}$. A numerical value of $M_{\rm
min}$ depends strongly on the nature of DM particle. Even in the
case of a particular DM particle, e.g., neutralino, the calculated
value of $M_{\rm min}$ can differ by many orders of magnitude for
different sets of parameters in mSUGRA model, see the
Appendix~\ref{subsmmin}. The larger scale clumps are formed later.
The larger scale clumps host the smaller ones and are hosted
themselves by next larger clumps. Major parts of small-scale
clumps are destroyed by the tidal gravitational fields of their
host clumps. At small-mass scales, the hierarchial clustering is a
fast and complicated nonlinear process. The formation of new
clumps and their capturing by the larger ones are nearly
simultaneous processes because  at small-scales an effective index
of the density perturbation power spectrum is very close to a
critical value, $n\to -3$. The DM clumps are not totally
virialized when they are captured by hosts. The adiabatic
invariants cannot prevent the survival of cores at this stage
because there are not enough time for the formation of the
singular density profiles in clumps (an internal dynamical time of
clump is of the same order as its capture time by a host). We use
a simplified model to take into account the most important
features of hierarchial clustering.

In the model of spherical collapse (see for example \cite{cole}),
a formation time for a clump with an internal density $\rho$ is
$t=(\kappa \rho_{\rm eq}/\rho)^{1/2}t_{\rm eq}$, where
$\kappa=18\pi^2$, $\rho_{\rm eq}=\rho_0(1+z_{\rm eq})^3$ is a
cosmological density at the time of matter-radiation equality
$t_{\rm eq}$, $1+z_{\text{eq}}= 2.35\times10^4\Omega_mh^2$, and
$\rho_0= 1.9\times10^{-29}\Omega_mh^2\mbox{ g cm}^{-3}$. The index
``eq'' refers to quantities at the time of matter-radiation
equality $t_{\rm eq}$. The DM clumps of mass $M$ can be formed
from density fluctuations of a different peak-height
$\nu=\delta_{\rm eq}/\sigma_{\rm eq}(M)$, where $\sigma_{\rm
eq}(M)$ is a fluctuation dispersion on a mass-scale $M$ at the
time $t_{\rm eq}$. A mean internal density of the clump $\rho$ is
fixed at the time of the clump formation and according to
\cite{cole} is $\rho= \kappa\rho_{\rm eq}[\nu\sigma_{\rm
eq}(M)/\delta_c]^3$, where
$\delta_c=3(12\pi)^{2/3}/20\simeq1.686$.

A tidal destruction of clumps is a complicated process and depends
on many factors: the formation history of clumps, host density
profile, the existence of another substructures inside the host,
orbital parameters of individual clumps in the hosts, etc. Only in
numerical simulations, all these factors can be taken into account
properly. The first such simulation in the small-scale region was
produced in \cite{DieMooSta05}. We use a simplified analytical
approach by parameterizing the energy gains in tidal interactions
by the number of tidal shocks per dynamical time in the hosts.
Using the model \cite{gnedin1} for tidal heating, we determine the
survival time $T$, i.e. time of tidal destruction, for a chosen
small-scale clump due to the tidal heating inside of a host clump
with larger mass. During the dynamical time $t_{\rm
dyn}\simeq0.5(G\rho_h)^{-1/2}$, where $\rho_h$ is a mean internal
density of the host, the chosen small-scale clump may belong to
several successively destructed hosts. A clump trajectory in the
host experiences successive turns accompanied by the ``tidal
shocks'' \cite{gnedin1}. Similar shocks come from interactions
with other substructure, and in general due to any varying
gravitational field. For the considered small-scale clump with a
mass $M$ and radius $R$, the corresponding internal energy
increase after a single tidal shock is
\begin{equation}
 \Delta E\simeq\frac{4\pi}{3}\,\gamma_1G\rho_hMR^2,
 \label{dele}
\end{equation}
where a numerical factor $\gamma_1\sim1$. Let us denote by
$\gamma_2$ the number of tidal shocks per dynamical time $t_{\rm
dyn}$. The corresponding rate of internal energy growth for a
clump is $\dot E=\gamma_2\Delta E/t_{\rm dyn}$. A clump is
destroyed in the host if its internal energy increase due to tidal
shocks exceeds a total energy $|E|\simeq GM^2/2R$. As a result,
for a typical time $T=T(\rho,\rho_h)$ of the tidal destruction of
a small-scale clump with density $\rho$ inside a more massive host
with a density $\rho_h$ we obtain
\begin{equation}
 T^{-1}(\rho,\rho_h)=\frac{\dot E}{|E|}\simeq4\gamma_1\gamma_2
 G^{1/2}\rho_h^{3/2}\rho^{-1}.
\end{equation}
It turns out that a resulting mass function of small-scale clumps
(see this Sec. below) depends rather weakly on the value of the
product $\gamma_1\gamma_2$.

During its lifetime, a small-scale clump can stay in many host
clumps of larger mass. After tidal disruption of the first
lightest host, a small-scale clump becomes a constituent part of a
larger one, etc. The process of hierarchical transition of a
small-scale clump from one host to another occurs almost
continuously in time up to the final host formation, where the
tidal interaction becomes inefficient. The probability of clump
survival, determined as a fraction of the clumps with mass $M$
surviving the tidal destruction in hierarchical clustering, is
given by the exponential function $e^{-J}$ with
\begin{equation}
 J\simeq\sum\limits_{h} \frac{\Delta t_h}{T(\rho,\rho_h)}.
 \label{jsum}
\end{equation}
Here, $\Delta t_h$ is a difference of formation times $t_h$ for
two successive hosts, and summation goes over all clumps of
intermediate mass-scales, which successively host the considered
small-scale clump of a mass $M$. Changing the summation by
integration in (\ref{jsum}) we obtain
\begin{equation}
 J(\rho,\rho_f)=\int\limits_{t_1}^{t_f}\!\frac{dt_h}{T(\rho,\rho_h)}
 \simeq\gamma\frac{\rho_1-\rho_f}{\rho}
 \simeq\gamma\,\frac{\rho_1}{\rho}\simeq\gamma\,\frac{t^2}{t_1^2}\,,
 \label{sumint1}
\end{equation}
where
\begin{equation}
 \gamma=2\gamma_1\gamma_2\kappa^{1/2}G^{1/2}
 \rho_{\rm eq}^{1/2}t_{\rm eq}\simeq14(\gamma_1\gamma_2/3),
 \label{bigj14}
\end{equation}
and $t$, $t_1$, $t_f$, $\rho$, $\rho_1$ and $\rho_f$ are,
respectively, the formation times and internal densities of the
considered clump and of its first and final hosts. One may see
from Eq.~(\ref{sumint1}) that the first host provides a major
contribution to the tidal destruction of the considered
small-scale clump, especially if the first host density $\rho_1$
is close to $\rho$, and consequently $e^{-J}\ll 1$. Therefore,
Eq.~(\ref{sumint1}) gives a qualitatively correct description of
the tidal destruction. However, in the more detailed approach the
dependence of $\gamma$ on another parameters is possible to take
into account. As reasonable estimate, we will use the ansatz given
by Eq.~(\ref{sumint1}) for further calculation of mass function.

Now we need to track the number of clumps $M$ (originated from the
density peak $\nu$) which enter some larger host during time
intervals $\Delta t_1$ around each $t_1$ beginning from the time
$t$ of clump formation. A mass function of small-scale clumps
(i.e., a differential mass fraction of DM in the form of clumps
survived in hierarchical clustering) can be expressed as
\begin{eqnarray}
\label{phiin}
 &&\xi\frac{dM}{M}\,d\nu= \\ &&
 dM\,d\nu\,\frac{e^{-\nu^2/2}}{\sqrt{2\pi}}\!\!
 \int\limits_{t(\nu\sigma_{\rm eq})}^{t_0}\!\!\!dt_1
 \left|\frac{\partial^2 F(M,t_1)}{\partial M~\partial t_1}\right|
 e^{-J(t,t_1)}. \nonumber
\end{eqnarray}
In this expression, $t_0$ is the Universe age and $F(M,t)$ is a
mass fraction of unconfined clumps (i.e., clumps, not belonging to
more massive hosts) with a mass smaller than $M$ at time $t$.
According to \cite{cole}, the mass fraction of unconfined clumps
is $F(M,t)={\rm erf}\left( \delta_c/[\sqrt{2}\sigma_{\rm
eq}(M)D(t)]\right)$, where ${\rm erf}(x)$ is the error-function
and $D(t)$ is the growth factor normalized by $D(t_{\rm eq})=1$.
An upper limit of integration $t_0$ in Eq.~(\ref{phiin}) is not
crucial and may be extrapolated to infinity because a main
contribution to the tidal destruction of clumps is provided by the
early formed hosts at the beginning of the hierarchical
clustering.

Two processes are responsible for time evolution of the fraction
$\partial^2 F/(\partial M\partial t)$ for unconfined clumps in the
mass interval $dM$: (i) the formation of new clumps and (ii) the
capture of smaller clumps into the larger ones. Both these
processes are equally efficient at the time when $\partial^2
F/(\partial M\partial t)=0$. To take into account the confined
clumps (i.e., clumps in the hosts) we need only the second process
(ii) for the fraction $\partial F(M,t)/\partial M$. Nevertheless,
in Eq.~(\ref{phiin}), which  it is used the fraction $\partial
F(M,t)/\partial M$ which depends on both processes. This is not
accurate at a typical formation time of a clump with a mass $M$,
when clump density is comparable with the density of hosts.
Fortunately, for this time the exponent in Eq.~(\ref{phiin}) is
very small, $e^{-J}\ll 1$, as it can be seen from (\ref{sumint1})
and (\ref{bigj14}). Respectively, an uncertain contribution from
the process (i) to the integral (\ref{phiin}) is also very small.
Meanwhile, only process (ii) dominates in the integration region
where the exponent $e^{-J}$ is not small. For this reason,
Eq.~(\ref{phiin}) provides a suitable approximation for the mass
fraction of clumps survived in the hierarchical clustering. The
characteristic epoch $t_*$ of the clumps $M$ formation can be
estimated from the equation $\sigma_{\rm
eq}(M)D(t_*)\simeq\delta_c$. If one considers, the times $t\gg
t_*$, then the exponents $\exp\!\left\{-\delta_c^2/[2\sigma_{\rm
eq}^2D^2(t)]\right\}$ can be putted approximately to unity for
simplification of integration in (\ref{phiin}).

\begin{figure}[t]
\includegraphics[width=0.45\textwidth]{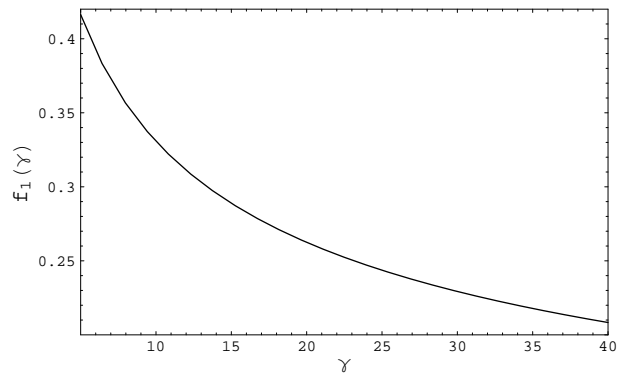}
\caption{The function $f_1(\gamma)$ from (\ref{f1fun}).}
 \label{fun1}
\end{figure}

Finally, we transform the distribution function (\ref{phiin}) to
the following form:
\begin{equation}
 \xi\,\frac{dM}{M}\,d\nu\simeq
 \frac{\nu\,d\nu}{\sqrt{2\pi}}\,e^{-\nu^2/2}
 f_1(\gamma)\frac{d\log\sigma_{\rm eq}(M)}{dM}\,dM,
 \label{psiitog}
\end{equation}
where
\begin{equation}
 f_1(\gamma)=
 \frac{2[\Gamma(1/3)-\Gamma(1/3,\gamma)]}{3\sqrt{2\pi}\gamma^{1/3}},
 \label{f1fun}
\end{equation}
$\Gamma(1/3)$ and $\Gamma(1/3,\gamma)$ are the Euler
gamma-function and incomplete gamma-function, respectively. The
function (\ref{f1fun}) is shown in Fig.~\ref{fun1}. It is seen in
this figure that $f_1(\gamma)$ varies rather slowly in the
interesting interval of $14<\gamma<40$, and one may use
$f_1(\gamma)\simeq0.2-0.3$.

Physically, the first factor $\nu$ in (\ref{psiitog}) corresponds
to a more effective survival of high-density clumps (i.e., with
large values of $\nu$) with respect to the low-density ones (with
small values of $\nu$). Integrating Eq.~(\ref{psiitog}) over
$\nu$, we obtain
\begin{equation}
 \xi_{\rm int}\frac{dM}{M}\simeq0.02(n+3)\,\frac{dM}{M}.
 \label{xitot}
\end{equation}
An effective power-law index $n$ in Eq.~(\ref{xitot}) is
determined as $n=-3(1+2\partial\log\sigma_{\rm eq}(M)/\partial\log
M)$ and depends very weakly on $M$. Equation (\ref{xitot}) implies
that for suitable values of $n$ only a small fraction of clumps,
about $0.1-0.5$~\%, survives the stage of hierarchical tidal
destruction in the each logarithmic mass interval $\Delta\log
M\sim1$. It must be stressed that a physical meaning of the
survived clump distribution function $\xi\,dM/M$ is different from
the similar one for the {\em unconfined} clumps, given by the
Press-Schechter mass function $\partial F/\partial M$.

The simple $M^{-1}$ shape of the mass function (\ref{xitot}) is in
very good agreement with the corresponding numerical simulations
\cite{DieMooSta05}, but our normalization factor is a few times
smaller.  One also can see a reasonable agreement between the
extrapolation of our calculations and the corresponding numerical
simulations of the large-scale clumps with $M\ge10^6M_\odot$ (for
a comparison see \cite{bde06}). The obtained mass function
(\ref{xitot}) is further transformed in the process of tidal
destructions of clumps by stars in the Galaxy (see previous
sections).

\section{Modified distribution function of clumps}
\label{modif}

In this section we calculate the modified mass function for the
small-scale clumps in the Galaxy taking into account clump mass
loss instead of the clump destruction considered in
\cite{bde03,bde07}.

\begin{figure}[t]
\includegraphics[width=0.45\textwidth]{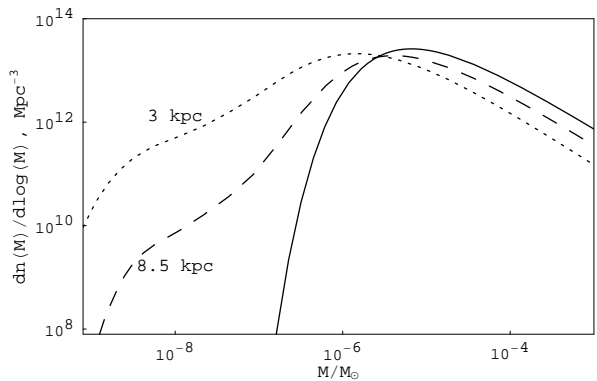}
\caption{Numerically calculated modified mass function of clump
remnants for galactocentric distances $3$ and $8.5$~ kpc. The
solid curve shows the initial mass function.}
 \label{mf1}
\end{figure}

According to theoretical model \cite{bde03} and  numerical
simulations \cite{DieMooSta05}, a differential number density of
small-scale clumps  in the comoving frame in the Universe is
$n(M)\,dM \propto dM/M^2$. This distribution is shown in
Fig.~\ref{mf1} by the solid line. The damping of small-scale
perturbations with $M<M_{\rm min}$ provides an additional factor
$\exp[-(M/M_{\min})^{2/3}]$ responsible for the fading of
distribution at small $M$. The result of the numerical simulations
\cite{DieMooSta05} can be expressed in the form of a differential
mass fraction of the DM clumps in the Galactic halo
$f(M)\,dM\simeq \kappa(dM/M)$, where
$\kappa\simeq8.3\times10^{-3}$. The analytical estimation
(\ref{xitot}) gives approximately $\kappa\simeq4\times10^{-3}$ for
the mass interval $10^{-6}M_\odot<M<1M_\odot$. The discrepancy by
the factor $\simeq2$ may be attributed to the approximate nature
of our approach as well as to the well known additional factor 2
in the original Press-Schechter derivation of the mass function.
In the later case one must simply multiply equation (\ref{phiin})
by factor 2. To clarify this discrepancy the more sophisticated
calculations are necessary.

As it was described earlier, we consider DM clumps originated from
$2\sigma$ density peaks. Therefore, in our approximation the
density of clumps and their distribution depends only on one
parameter $M$. In general, the distribution of DM clumps depends
on the pair of parameters, e.g. mass and radius, mass and velocity
dispersion or mass and peak-high $\nu$ as in the distribution
(\ref{psiitog}). Meanwhile the authors of numerical simulations do
not present a general distribution of clumps over two parameters.
The general distribution of clumps can be in principle extracted
from simulations and is very requested for further investigations
of DM clumpiness.

By using the formalism of Sec.~\ref{distrib}, we derive the mass
distribution of the clump remnants in dependence of the initial
masses $M_i$ of clumps. To do this, we calculate numerically the
value of the mass $\mu$ of the clump remnant in dependence of the
initial mass $M_i$ for separate elements $\Delta
p\Delta\gamma\Delta\varepsilon$ in the parameter space in
(\ref{sp1}). Then for fixed intervals $\Delta\mu$ of values of
$\mu$, we provide the summation of the weights of distribution
function, which is given by Eq.~(\ref{sp1}) without symbols of
integration and $\Theta$-function. By using the derived
$\mu$-distribution, we transform the initial (cosmological) mass
function of clumps to the final (nowadays) mass function in the
halo at the present moment. This final mass function is shown in
Fig.~\ref{mf1} for two distances from the Galactic center. We
supposed in numerical calculations that a core radius is very
small and all masses of remnants are admissible. With a finite
core size, the final mass function has a cutoff near the cores
mass of clump with a minimal mass $M_{\rm min}$.  The adiabatic
correction leads to the accumulation of remnants of some mass
corresponding to violation of momentum approximation. One can see
from Fig.~\ref{mf1} that clump remnants exist below the $M_{\rm
min}$. Deep in the bulge (very near to the Galactic center) the
clump remnants are more numerous because of intensive destructions
of clumps in the dense stellar environment in comparison with the
rarefied one in the halo. The main contribution to the low-mass
tail of the mass function of remnants comes from the clumps with
the near-disk orbits where the destructions are more efficient.

The another important point is an efficient destruction of clumps
with orbits confined inside the stellar bulge. Nevertheless, a
number density of clumps inside the bulge is nonzero because a
major part of clumps have orbits extending far beyond the bulge,
These ``transit'' clumps spend only a small part of their orbital
time traversing the bulge and survive the tidal destruction.

\section{Amplification of annihilation signal}

A local annihilation rate is proportional to the square of the DM
particle number density. A number density of DM particles in clump
is much large than a corresponding number density of the diffuse
(not clumped) component of DM. For this reason, an annihilation
signal from even a small fraction of DM clumps can dominate over
an annihilation signal from the diffuse component of DM in the
halo. In this section, we calculate the amplification (or
``boosting'') of an annihilation signal due to the presence of the
survived DM clump remnants in the Galactic halo. We consider here
the Harrison-Zeldovich initial perturbation spectrum with power
index $n_p=1$ as a representative  example. The value of $n_p$ is
not exactly fixed by the current observations of cosmic microwave
background (CMB) anisotropy. In the case of $n_p<1$, the DM clumps
are less dense, and a corresponding amplification of annihilation
signal would be rather small \cite{bde03}.

\begin{figure}[t]
\includegraphics[width=0.45\textwidth]{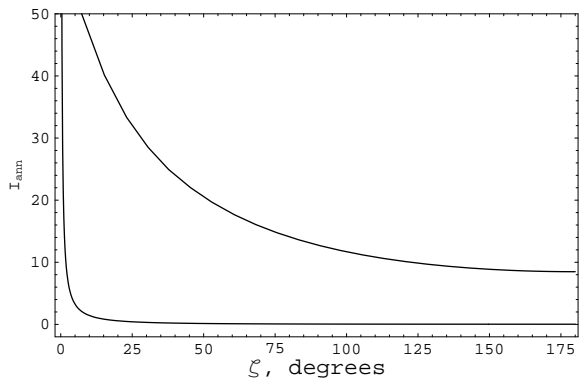}
\caption{The annihilation signal (\ref{ihal2}) (upper curve) as a
function of the angle $\zeta$ between the line of observation and
the direction to the Galactic center. For comparison the
annihilation signal is also shown (by the bottom curve) from the
Galactic halo without DM clumps (\ref{ihal1}). The values of both
integrals (\ref{ihal2}) and (\ref{ihal1}) are multiplied by a
factor of $10^{48}$.} \label{boo}
\end{figure}

\begin{figure}[t]
\includegraphics[width=0.45\textwidth]{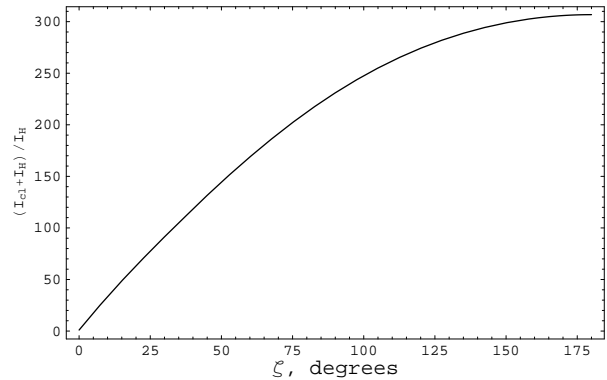}
\caption{The amplification of the annihilation signal $(I_{\rm
cl}+I_{\rm H})/I_{\rm H}$ as function of the angle between the
line of observation and the direction to the Galactic center,
where fluxes are given by (\ref{ihal1}) and (\ref{ihal2}).}
\label{boo2}
\end{figure}

The gamma-ray flux from the annihilation of the diffuse
distribution (\ref{halonfw}) of DM in the halo is proportional to
\begin{equation}
 I_{\rm H}=\int\limits_{0}^{r_{\rm max}(\zeta)}\rho_{H}^2(\xi)\,dx,
 \label{ihal1}
\end{equation}
where the integration is over $r$ goes along the line of sight,
$\xi(\zeta,r) = (r^2+r_{\odot}^2-2rr_{\odot}\cos\zeta)^{1/2}$ is
the distance to the Galactic center, $r_{\rm max}(\zeta) = (R_{\rm
H}^2-r_{\odot}^2\sin^2\zeta)^{1/2} + r_{\odot}\cos\zeta$  is a
distance to the external halo border, $\zeta$ is an angle between
the line of observation and the direction to the Galactic center,
$R_{\rm H}$ is a virial radius of the Galactic halo,
$r_{\odot}=8.5$~kpc is the distance between the Sun and the
Galactic center. The corresponding signal from annihilations of DM
in clumps is proportional to the quantity \cite{bde03}
\begin{equation}
 I_{\rm cl}= S\int\limits_{0}^{r_{\rm max}(\zeta)}\!dx\!\!
 \int\limits_{M_{\rm min}}\!\!f(M)\,dM
 \rho\rho_{H}(\xi)P(\xi,\rho),
 \label{ihal2}
\end{equation}
where $\rho(M)$ is the mean density of the clump. The function $S$
depends on the clump density profile and core radius of the clump
\cite{bde03}, and we use $S\simeq 14.5$ as a representative
example. The observed amplification of the annihilation signal is
defined as $\eta(\zeta)=(I_{\rm cl}+I_{\rm H})/I_{\rm H}$ is shown
in Fig.~\ref{boo2} for the case $x_c=0.1$. It tends to unity at
$\zeta\to 0$ because of the divergent form of the halo profile
(\ref{halonfw}). The annihilation of diffuse DM prevails over
signal from clumps at the the Galactic center. The $\eta(\zeta)$
very slightly depends on $x_c$, and corresponding graphs for
$x_c<0.1$ are almost indistinguishable from the one in
Fig.~\ref{boo}. This is because the observed signal is obtained by
integration along the line of sight and the effect of the clump's
destruction at the Galactic center is masked by the signal from
another regions of the halo.

This amplification of an annihilation signal is often called a
``boost-factor''. A boost-factor of the order of $10$ is required
for interpretation of the observed EGRET gamma-ray excess as a
possible signature of DM neutralino annihilation \cite{deBoer}.

\section{Conclusion}

In \cite{ZTSH} it was found that almost all small-scale clumps in
the Galaxy are destructed by tidal interactions with stars and
transformed into ``ministreams'' of DM. The properties of these
ministreams may be important for the direct detection of DM
particles because DM particles in streams arrive anisotropically
from several discrete directions. In this work, we demonstrate
that the cores of clumps (or clump remnants) survive  in general
during the tidal destruction by stars in the Galaxy. Although
their outer shells are stripped and produce the ministreams of DM,
the central cores are protected by the adiabatic invariant and
survived as the sources of annihilation signals. This conclusion
depends crucially on the unknown sizes of the cores: the smaller
cores are more protected because DM particles there have higher
orbital frequencies and therefore the larger the adiabatic
parameter.

Despite the  small survival probability of clumps during early
stage of hierarchial clustering, they provide the major
contribution to the annihilation signal (in comparison with the
unclumpy DM). The amplification (boost-factor) can reach $10^2$ or
even $10^3$ depending on the initial perturbation spectrum and
minimum mass of clumps. This boost-factor must be included in
calculations of the annihilation signals. Some promising
interpretations of observations and calculations of annihilation
signal from the Galactic halo require this boost-factor (see,
e.g., \cite{deBoer}). The discussed dense remnants survive the
tidal destruction and provide the enhancement of DM annihilation
in the Galaxy. These remnants of DM clumps form the low-mass tail
in the standard mass distribution of small-scale clumps extended
much below $M_{\rm min}$ of the standard distribution. It does not
mean of course the increasing of annihilation signal in comparison
with the case without clump destruction. It only indicates that
galactic clump destruction does not diminishes strongly the
annihilation signal.

The principle simplifying assumption of this work is that only the
outer layers of clumps are subjected by the tidal stripping. The
main difficulties in considering the full problem with the mass
loss from inner layers are in the complicated dynamical
reconstruction of clumps just after tidal shocks. We believe that
our approach provides a rather good result by the two reasons.
First, the influence of tidal forces depends on the system size,
and, therefore, the outer layers are greatly subjected to tidal
forces. Second, the adiabatic protection is more efficient in the
inner part of the clump because of higher orbital frequencies
here. In reality, we expect some expansion of clump and
diminishing of its central density due to energy deposit from
tidal forces. It is very interesting task to clarify this process
in future works.

The numerical estimate of the boost-factor for DM particle
annihilation inside clumps is very model-dependent. It depends on
nature of DM particles and on their interaction with ambient
plasma. The important physical parameters, which affects the
annihilation rate in clumps, are decoupling temperature $T_d$ and
minimal mass $M_{\rm min}$ in the clump mass distribution. The
boost-factor increases strongly for small $M_{\rm min}$. The
minimal mass in standard calculations is determined by the escape
of DM particles from a growing fluctuation due to, e.g.,
diffusion, free streaming or Silk effect. Uncertainties in the
calculated values of $T_d$ and $M_{\rm min}$ are discussed in
Appendix \ref{subsmmin}. For the lightest neutralino as a DM
particle, assuming it to be the pure bino $B$, one can see from
Table~III a huge difference in $M_{\rm min}$ caused by the
variation of supersymmetry (SUSY) parameters $m_{\chi}$ and
$\tilde{m}$. For these parameters, we use cosmologically allowed
values from the benchmark scenarios of the work \cite{ellis}.
Moreover, inclusion of other neutralino compositions, e.g., mixed
bino-Higgsino, the other allowed benchmark scenarios with
co-annihilation and focus-point regions, and some other
modifications, may very considerably increase the allowed  region
of $M_{\rm min}$ values up to $(3\times 10^{-12} - 7\times
10^{-4})M_{\odot}$ \cite{Kam06}. Inclusion of the other particle
candidates extends further this region.

Another parameter variation which affects strongly the
boost-factor is the spectral index of density perturbation $n_p$
(see \cite{bde03}). We conclude thus that the annihilation
boost-factor (enhancement) even for neutralino has large
uncertainties due to the difference in SUSY papameters and
spectral index $n_p$. It can reach the factor $10^4$ and even
more, The largest values of boost-factor can be already excluded
by observations of indirect signal, since mSUGRA parameters can be
fixed for this largest value. On the contrary, a tidal destruction
of clumps in the Galaxy affects the annihilation boost-factor much
weaker.

\begin{acknowledgments}
We thank K. P. Zybin and V. N. Lukash for helpful discussions.
This work has been supported in part by the Russian Foundation for
Basic Research grants 06-02-16029 and 06-02-16342, the Russian
Ministry of Science grants LSS 4407.2006.2 and LSS 5573.2006.2.
\end{acknowledgments}

\appendix

\section{Minimum clump mass}

\subsection{Uncertainties in minimal clump mass and decoupling
temperature}
 \label{subsmmin}

The low-mass cutoff of the clump mass-spectrum accompanies the
process of decoupling. It starts when DM particles  coupled
strongly with surrounding plasma in the growing density
fluctuations. The smearing of the small-scale fluctuations is due
to the collision damping occurring just before decoupling, in
analogy with the Silk damping \cite{Silk68}.  It occurs due to the
diffusion of DM particles from a growing fluctuation, and only the
small-scale fluctuations can be destroyed by this process. The
corresponding diffusive cutoff $M_{\rm min}^{\rm diff}$ is very
small. As coupling becomes weaker, the larger fluctuations are
destroyed and $M_{\rm min}$ increases. One may expect that the
largest value of $M_{\rm min}$ is related to a free-streaming
regime. However, as recent calculations show \cite{bertsh}, the
largest $M_{\rm min}$ is related to some friction between DM
particles and cosmic plasma similar to the Silk damping. The
predicted minimal clump masses range from very low values, $M_{\rm
min}\sim10^{-12}M_\odot$ \cite{zvg}, produced by diffusive escape
of DM particles, up to $M_{\rm min}\sim10^{-4}M_\odot$, caused by
acoustics oscillations \cite{loebz} and quasi-free-streaming with
limited friction \cite{bertsh}.

The calculations of minimal clump mass $M_{\rm min}$ and
decoupling temperature $T_{\rm d}$ are determined by elastic
scattering of DM particles off leptons $l=(\nu_{L},e_{L},e_{R})$
in cosmic plasma. The uncertainties in cross-section very strongly
influence the resulting values of $M_{\rm min}$ and $T_{\rm d}$.
In all works cited above, the lightest neutralino ($\chi$) in the
form of pure bino ($\tilde B$) is assumed as a DM particle, and
$\chi l$-scattering occurs due to exchange by sleptons $\tilde
l=(\tilde\nu_{L},\tilde e_{L},\tilde e_{R})$.

The elastic cross-sections for $l_{L}\chi$ and $l_{R}\chi$
scattering have been calculated in \cite{bde03} as
\begin{equation}
 \left(\frac{d\sigma}{d\Omega}\right)_{l_L\chi}\!=
 \frac{\alpha_{\text{em}}^2}{8\cos^4\theta_{\rm W}}
 \frac{\omega^2(1+\cos\theta_{\rm cm})}{(\tilde m_L^2-m_{\chi}^2)^2}
 \label{crosss1}
\end{equation}
and
\begin{equation}
 \left(\frac{d\sigma}{d\Omega}\right)_{l_R\chi}\!=
 16\left(\frac{d\sigma}{d\Omega}\right)_{l_L\chi}\!,
 \quad \mbox{if} \quad  \tilde m_L=\tilde m_R,
 \label{crosss2}
\end{equation}
where $\omega\gg m_l$ is a c.m.-energy of $l$, $\theta_{\rm cm}$
is a scattering angle of $l$ in c.m.-system, $m_{\chi}$ is a
neutralino (bino) mass, $\tilde m_L$ and $\tilde m_R$ are,
respectively a mass of the left and right sfermions, and
$\theta_{\rm W}$ is the Weinberg angle.

The values of $T_{\rm d}$ and $M_{\rm min}$ as cited in
\cite{bde03,shs,GreHofSch05,loebz} differ very much from each
other, but a very big contribution to this difference comes from
the differences in the used values for $m_{\chi}$ and $\tilde m$.
To see the difference, which must be attributed to the different
damping mechanisms used in these works, we recalculated $T_{\rm
d}$ and $M_{\rm min}$ with the same values of $m_{\chi}$ and
$\tilde m$, for which we used $100$ and $200$~GeV respectively.
The results are presented in Table~I. We did not include there the
work \cite{zvg} because a pure diffusive damping results in too
low a value of $M_{\rm min}$.
\begin{table}
 \label{tab1}
\begin{tabular}{ccccccc}
 \hline
& Reference: &\cite{shs}$^{1)}$ & \cite{bde03}$^{1)}$ &
\cite{GreHofSch05}$^{2)}$
& \cite{loebz}$^{3)}$& \cite{bertsh}$^{4)}$ \\
 \hline
&$T_d$,~MeV & 28 & 26 & 25 & 20 & 22.6  \\
&$M_{\rm min}/M_\odot$ & $2.5\times10^{-7}$ & $1.7\times10^{-7}$
& $1.5\times10^{-6}$ & $1.3\times10^{-5}$& $8.4\times10^{-6}$\\
\end{tabular}
\caption{The values of decoupling temperature $T_{\rm d}$ and
minimal clump mass $M_{\rm min}$ with $m_{\chi}=100$~GeV and
$\tilde m=200$~GeV for different damping mechanisms:
$^{1)}$free-streaming, $^{2)}$collision damping, $^{3)}$acoustic
oscillations, $^{4)}$quasi-free-streaming with friction.}
\end{table}
\begin{table}
 \label{tab2}
\begin{tabular}{cccccccccccc}
 \hline
scenario && $\chi$ && $\tilde e_L$ && $\tilde e_R$ &&
$\tilde\nu_e,\tilde\nu_\mu$ && $\tilde\nu_\tau$ \\
 \hline
B' && 95 && 188 && 117 && 167 && 167 \\
E' && 112 && 1543 && 1534 && 1539 && 1532 \\
M' && 794 && 1660 && 1312 && 1648 && 1492 \\
\end{tabular}
\caption{Selected benchmark scenarios from \cite{ellis}. The
masses of particles are given in GeV.}
\end{table}
From the Table~I one can see a reasonable agreement in values of
$T_{\rm d}$ and $M_{\rm min}$, and a successive increasing of
$M_{\rm min}$ from $2.5\times 10^{-7}M_\odot$ for free-streaming
to $\sim10^{-5}M_\odot$ for oscillation damping and
quasi-free-streaming with friction.

\subsection{Uncertainties in SUSY parameters}
\label{bench}

We shall consider now the range of predictions for different
values of SUSY parameters allowed in cosmology. For this aim, we
shall use the SUSY benchmark scenarios from the work \cite{ellis},
which agree with the Wilkinson Microwave Anisotropy Probe (WMAP)
and other cosmological data. These benchmark scenarios are
obtained within mSUGRA model with universal parameters at the
grand unified theory scale: $m_0$ (the universal scalar soft
breaking mass), $m_{1/2}$ (the universal gaugino soft breaking
mass), $A_0$ (the universal cubic soft breaking terms) and
$\tan\beta$ (the ratio of two Higgs v.e.v.'s). The LEP and $b\to
s\gamma$ constraints are imposed. The resulting relic density of
neutralinos from these scenarios is in agreement with the WMAP
data or can be obtained with small changes of $m_0$ and $m_{1/2}$.
In Table~II we display three benchmark scenarios from
\cite{ellis}. The scenario B' gives the lower value
$m_\chi\approx100$~GeV and $\tilde m$ close to $200$~GeV, which we
discussed above. Scenario M' gives the highest value
$m_\chi\approx800$~GeV and $\tilde m\approx1600$~GeV.
Respectively, scenario E' gives the intermediate value
$m_\chi\approx110$~GeV and $\tilde m\approx1500$~GeV, similar to
those we used in \cite{bde03}.

To illustrate the uncertainties in $T_{\rm d}$ and $M_{\rm min}$
due to uncertainties in $m_\chi$ and $\tilde m$  (in the
simplifying assumption that $m_{\tilde\nu}=m_{\tilde
e_L}=m_{\tilde e_R}$) we choose the calculations of Bertschinger
\cite{bertsh} in quasi-free-streaming scenario with friction,
which seem to be at present the most detailed ones. We use the
Bertschinger formulaes
\begin{equation}
 \label{kindec}
  T_d=7.65\,C^{-1/4}g_{*}^{1/8}\left(\frac{m_\chi}{\hbox{100
    GeV}}\right)^{5/4}\ \hbox{MeV},
\end{equation}
\begin{equation}
 \label{mmin}
 M_{\rm min}=
 7.59\times10^{-3}\,C^{3/4}\!\left(\frac{m_\chi\sqrt{g_{*}}}
    {\hbox{100 GeV}}\right)^{-15/4}\,M_\odot,
\end{equation}
with a dimensionless constant
\begin{equation}
 \label{Cneut}
  C=256\,(G_Fm_W^2)^2 \left(\frac{\tilde m^2}{m_\chi^2}
  -1\right)^{-2} \sum_{L}(b_L^4+c_L^4),
\end{equation}
where $G_{\rm F}$ is the Fermi coupling constant, $b_L$ and $c_L$
are left and right chiral vertices, and $m_W$, $\tilde m$, and
$m_\chi$ are, respectively, the masses of the $W$ boson
$G_Fm_W^2=0.0754$, the slepton, the neutralino and the number of
freedom at the decoupling epoch $g_*=43/4$. (Our own calculations
in \cite{bde03} of $C$, which is related with the square of the
matrix element for $l+\chi\to l+\chi$ scattering, differ from
(\ref{mmin}) by a factor $1.6$.) As a result we obtain for the
benchmark scenarios which approximately coincide with model B'
(minimum $m_\chi$ and $\tilde m$), E' (minimum $m_\chi$ and large
$\tilde m$) and M' (very large $m_\chi$ and $\tilde m$) the values
of $T_{\rm d}$ and $M_{\rm min}$ listed in Table~III.
\begin{table}
 \label{tab3}
\begin{tabular}{ccccccc}
 \hline
$m_\chi$ && $\tilde m$ && $T_{\rm d}$ && $M_{\rm min}$ \\
 \hline
$100$~GeV && $200$~GeV && $22.6$~MeV && $8.4\times10^{-6}M_\odot$ \\
$100$~GeV && $1500$~GeV && $196$~MeV && $1.3\times10^{-8}M_\odot$ \\
$800$~GeV && $1600$~GeV && $305$~MeV && $3.5\times10^{-9}M_\odot$ \\
\end{tabular}
\caption{Values of $T_{\rm d}$ and $M_{\rm min}$ for the
Bertschinger \cite{bertsh} damping scenario and three benchmark
scenarios \cite{ellis} which close to scenarios B', E' and M'
shown in the Table~II.}
\end{table}
The predicted range of parameters for $M_{\rm min}$ from this
Table:\ $3.5\times 10^{-9}-8.4\times 10^{-6}M_\odot$ is not robust
at all. It is obtained within mSUGRA assumptions about possible
universality of SUSY parameters $m_0$, $m_{1/2}$ and $A_0$.
Lifting the universality restriction, the mass of the neutralino
can increase up to the TeV range scale (though $m_\chi>200$~GeV
needs a fine-tuning less than $1$\% in SUSY \cite{bbe} or
decreased down to a few GeV \cite{bfs}).

In the numerical predictions above, we limited ourselves by rather
restrictive assumptions on the mSUGRA model. The most important of
them are assumption that the neutralino is a pure bino state and a
choice of cosmologically allowed benchmark scenario. The detailed
analysis made in \cite{Kam06} showed that allowed parameters of
mSUGRA result in much wider possibilities, e.g., neutralino as
mixed bino-Higgsino and the other benchmark scenarios. These
possibilities are considered under WMAP cosmological constraints
and a condition of producing the corresponding DM density for each
set of mSUGRA parameters. The considered modifications allow new
channels of neutralino interactions with ordinary particles, e.g.,
the exchange by $Z$-boson, co-annihilation and resonances in
neutralino-fermion scattering. It results in a wide range (many
orders of magnitude) of scattering cross-sections, and,
respectively, in a wide range of decoupling temperature from
$5$~MeV to $3$~GeV.  The corresponding range of $M_{\rm min}$ is
given by $(3\times 10^{-12} - 7\times 10^{-4})~M_{\odot}$. The
authors consider also the Kaluza-Klein particle as a DM candidate.

The small-scale mass of $M_{\rm min}$ results in the large density
of DM clumps, and thus in a much stronger annihilation signal from
the Galactic halo. For typical values of power-index of
perturbation spectrum (from CMB observations) the small-scale mass
of $M_{\rm min}$ results in the large density of DM clumps, and
thus in a much stronger annihilation signal from the Galactic
halo. However, the dependance of a mean clump density on the clump
mass is rather weak due to the nearly flat form of the
perturbation spectrum at small scales. The crucial factor for the
amplification of the annihilation signal by clumps is the value of
the perturbation power-law index $n_p$.

\end{document}